\documentclass[pof,reprint,aip,a4paper,twocolumn,amssymb, amsmath, superscriptaddress,citesort]{revtex4-1}
\usepackage{bm}%
\usepackage{epsfig}
\usepackage[colorlinks=true,linkcolor=blue]{hyperref}%
\expandafter\ifx\csname package@font\endcsname\relax\else
 \expandafter\expandafter
 \expandafter\usepackage
 \expandafter\expandafter
 \expandafter{\csname package@font\endcsname}%
\fi

\newcommand{\REM}[1]{}
\begin{document}
\title{Droplet size distribution in homogeneous isotropic turbulence}
\author{Prasad Perlekar}
\affiliation{International Collaboration for Turbulence Research,
  Dept.\ Physics and Dept.\ Mathematics \& Computer Science and
  J. M. Burgers Centre for Fluid Dynamics,\\Eindhoven University of
  Technology, 5600 MB Eindhoven, The Netherlands.}
\email[Corresponding author:]{p.perlekar@tue.nl}
\author{Luca Biferale}
\affiliation{International Collaboration for Turbulence Research,
  Dept. Physics and INFN, University of Rome ``Tor Vergata'' Via della Ricerca
  Scientifica 1, 00133 Rome, Italy.}
  \author{Mauro Sbragaglia}
\affiliation{International Collaboration for Turbulence Research,
  Dept. Physics and INFN, University of Rome ``Tor Vergata'' Via della Ricerca
  Scientifica 1, 00133 Rome, Italy.}
 \author{Sudhir Srivastava}
 \affiliation{Dept.\ Physics and Dept.\ Mathematics \& Computer Science and
  J. M. Burgers Centre for Fluid Dynamics,\\Eindhoven University of
  Technology, 5600 MB Eindhoven, The Netherlands.}
\author{Federico Toschi}  
\affiliation{International Collaboration for Turbulence Research,
  Dept.\ Physics and Dept.\ Mathematics \& Computer Science and
  J. M. Burgers Centre for Fluid Dynamics,\\Eindhoven University of
  Technology, 5600 MB Eindhoven, The Netherlands and\\ CNR-IAC, Via dei
  Taurini 19, 00185 Rome, Italy.}
\begin{abstract}
  We study the physics of droplet breakup in a statistically
  stationary homogeneous and isotropic turbulent flow by means of high
  resolution numerical investigations based on the multicomponent
  lattice Boltzmann method. We verified the validity of the criterion
  proposed by Hinze (1955) for droplet breakup and we measured the
  full probability distribution function (pdf) of droplets radii at
  different Reynolds numbers and for different volume fraction. By
  means of a Lagrangian tracking we could follow individual droplets
  along their trajectories, define a local Weber number based on the
  velocity gradients and study its cross-correlation with droplet
  deformation.
\end{abstract}
\date{\today}%

\maketitle
  \section{Introduction}
Droplets in turbulent flows occur in variety of industrial processes
such as sprays, colloid mills, and mixers.  When the droplet diameter
are smaller than the dissipative scales of turbulence, one can
approximate them as point particles whose dynamics is governed by the
Maxey-Riley-Gatignol equations \cite{mr83,gat83}. Several numerical,
analytical, and experimental studies have studied this regime and
found interesting phenomena such as clustering of particles at small
and large scales~\cite{tos09,bec07}. When the diameter of the
droplet is larger than the dissipative scales of turbulence, the point
particle approximation is no longer valid. Inertial scale sized
droplets deform, break, and coalesce under the action of turbulence.
The degree of deformation is governed by the ratio of the surface
tension forces and the intensity of turbulence or the Weber number,
$We$:
\begin{equation}
We \equiv \frac{\rho^{(m)} \langle(\delta u_D)^2\rangle D}{\sigma},
\end{equation} 
where $\rho^{(m)}$ is the density of the carrier medium fluid, $\delta
u_D$ is the average velocity difference across the droplet, the
angular bracket indicate spatial averaging, $D$ the droplet diameter,
and $\sigma$ the surface tension. If inertia forces overwhelm the
surface tension $We>1$, droplet breaks.  Using Kolmogorov theory for
velocity differences $\langle (\delta u_D)^2 \rangle\sim
\varepsilon^{2/3}D^{2/3}$ where $\varepsilon$ is the energy
dissipation rate, Hinze showed, in his 1955 seminal work \cite{hin55},
that the maximum droplet diameter that does not undergo breakup is
given by criterion:
\begin{equation}
  D_{max} = C \left(\frac{\rho^{(m)}}{\sigma}\right)^{-3/5}\varepsilon^{-2/5}
\label{eq:hinze}
\end{equation} 
where the coefficient $C=0.725$ is obtained by fitting with
experimental data. The above argument does not take into account the
coagulation of droplets which constitute an important mechanism in
dense suspensions. Even in dilute limit because of droplet breakup and
collision events one expects the droplet dispersion pdf to have a
finite width peaked around $D_{max}$. Breaking rate is strongly
correlated with the underlying turbulent stress {\it coarse grained}
on a scale of the size of droplet. The latter quantity in fully
developed turbulence is strongly non-Gaussian, with maximal deviations
from Gussianity observed in the viscous-inertial intermediate
range. Understanding the stationary droplet distribution due to
break-up and coagulation at changing Reynolds numbers and droplet
volume fraction remains a key unsolved problem.

Experimental studies by Pacek {\em et al.} \cite{pac94, pac98} focused
on turbulent dispersions where the carrier medium and the droplet
phase have the same density and viscosity and observed droplets size
distributions consistent with log-normal. Andersson {\em et al.}
\cite{and06} studied both bubble and droplet breakup experimentally
and found that single bubbles primarily undergoes binary breakups
while droplets typically present multiple breakups. Risso and Fabre
studied the oscillations and breakup of bubbles in turbulent
flow~\cite{fab98} while Ravelet {\em et al.}~\cite{rav11} focused on
breakups of a bubbles in turbulence rising due to gravity. The
experimental studies of Eastwood {\em et al.}~\cite{eas04} focused on
the breakup of bubbles in turbulent jets and reported deviations from
what predicted by the Hinze criterion~\cite{hin55}.

To date few numerical studies exist, this is probably due to the
additional difficulties implied with the need for interface tracking
under highly turbulent conditions.  Further, to reach good statistical
convergence, very long and computationally expensive stationary runs
are needed.  Numerical works by Qian {\em et al.}~\cite{qia06} studied
first few breakup events in a high-density contrast bubble breakup and 
were able to reproduce the experimental results of Fabre {\em et al.}~\cite{fab98}. 
Derksen {\em  et al.}~\cite{der07} studied how turbulence history can effect
coarsening by switching off turbulence in a droplet dispersion.

The aim of the present paper is to understand the interplay between
{\em turbulent fluctuations} and {\em surface tension} in a turbulent
emulsion. To this end, we always keep densities and viscosities of the
droplet and of the medium identical ($\rho^{(d)}/\rho^{(m)}=1$ and
$\nu^{(d)}/\nu^{(m)}=1$). We focus on cases with different droplet
volume fractions $\phi=V_d/V_m$, where $V_d$ and $V_m$ design volumes
of the droplet or of the carrier phase respectively. Furthermore, to
avoid any effects of boundaries, we limit to homogeneous and isotropic
turbulence. Our main results are: (a) we show that a stationary
droplet emulsion can be maintained for arbitrary long times, with a
chosen turbulence intensity and volume fraction, $\phi$; (b) the
average droplet diameter is in agreement with what predicted by the
Hinze criterion at least for small values of $\phi$; (c) droplet pdf
show peaks close to the average droplet radius and broadens at
increasing $\phi$; (d) by means of Lagrangian tracking of individual
droplets trajectory we show that breakup events are typically
associated with peaks in the local energy dissipation rate in the
droplet neighborhood.

{\section{Numerical method}}
We model the droplets and the carrier fluid phase by means of a
multicomponent Lattice-Boltzmann (LBM) algorithm with, by now
standard, non-ideal interactions as introduced by Shan-Chen
\citep{sha93,sha94,sha95}. This is a well established numerical
method thus we provide here only its key details.  The stirring
mechanism, which is needed in order to keep the system in a stationary
turbulent state, is applied via a large-scale forcing.

The lattice Boltzmann equations for the Shan-Chen multicomponent D3Q19
model are:
\begin{eqnarray}
  \nonumber
  f_{i}^{(\alpha)} ({\bm x} + {\bm e}_i,t +1) = f_{i}^{(\alpha)}({\bm x}) - \frac{1}{\tau_\alpha}[f_i^{(\alpha)}({\bm x},t ) -  \bar{f}^{(\alpha)}_i(\rho,{\bm u})] &\\
  \nonumber
  \bar{f}_i^{(\alpha)}(\rho,{\bm u}) = \rho^{(\alpha)} w_i \left[1+\frac{{\bm e}_i \cdot {\bm u}}{c_s^2} + \frac{{\bm u \bm u}:({\bm e}_i{\bm e_{i}} -c_s^2 {\bm I})}{2c_s^4}\right]&\\
  \nonumber
  \rho^{(\alpha)} =\sum_{i} f_i^{(\alpha)};~
  {\bm u}^{(\alpha)}({\bm x},t)=\sum_{i} {\bm e}_i f_{i}^{(\alpha)}({\bm x},t).&
\end{eqnarray}
Here $f_{i}^{\alpha}({\bm x},t)$ are the LBM distribution function at
position ${\bm x}$ and time $t$ for the fluid component
$\alpha=\{d,m\}$ (droplet and medium, respectively).
The fluid densities and velocities of the individual components are
$\rho^{(\alpha)}$ and ${\bm u}^{(\alpha)}$. Here $w_i$, ${\bm e}_{i}$
are the LBM weights and lattice speeds,
$i=\{0,\ldots,18\}$~\citep{succ01,Die00}. The total density of the
fluid is the sum of the density of the two components
$\rho=\sum_\alpha \rho^{(\alpha)}$ while the total hydrodynamic
velocity is ${\bm u}=\sum_{\alpha} u^{(\alpha)} \rho^{(\alpha)}/\rho$.
The global effective kinematic viscosity of the fluid is related to
the relaxation times of its components $\nu=\sum_{\alpha} c_s^2
(\tau^{(\alpha)} c^{(\alpha)}-0.5)$ \citep{sha95},
$c^{(\alpha)}=\rho^{(\alpha)}/\rho$ is the concentration, and
finally $c_s=1/\sqrt{3}$ is the speed of sound on the lattice for the D3Q19.

The non-ideal nature of the fluid is introduced by adding an extra
force to the LBM equilibrium velocity as~\citep{sha94}:
\begin{equation} 
{\bar{\bm u}}^{(\alpha)} = {\bm u}^{\prime} +
  \frac{\tau^{(\alpha)}{\bm F}^{(\alpha)}_{SC}}{\rho^{(\alpha)}}
  \;\;\;\mbox{and}\;\;\; {\bm u}^\prime=\frac{\displaystyle
    \sum_{\alpha} \rho^{(\alpha)} {\bm
      u}^{(\alpha)}/\tau^{(\alpha)}}{\displaystyle \sum_\alpha
    \rho^{(\alpha)}/\tau^{(\alpha)}}.
\end{equation}
The non-ideal interaction, as proposed by Shan and Chen, is~\citep{sha94}:
\begin{equation} {\bf F}^{(\alpha)}_{SC} = -G \rho^{(\alpha)} ({\bm
    x}) \sum_{i,\alpha\neq\alpha^\prime} \rho^{(\alpha^\prime)} ({\bm
    x} + {\bm e}_{i}) w_i {\bm e}_{i}
\end{equation}
where $\{\alpha,\alpha^{\prime}\}=\{d,m\}$ and the coupling parameter,
$G$, determines the strength of the microscopic interaction and
effectively sets the value of the surface tension and the
diffusivity~\citep{ben09sb}. Under appropriate conditions this force
allows the formation of interface between the different fluid
components.  

As already anticipated in this study we limit ourself to the case
$\tau^{(d)}=\tau^{(m)}=\tau$ which implies $\nu^{(d)}=\nu^{(m)}=\nu$,
i.e. identical viscosities for the two phases.  The total fluid
density $\rho^{(d)}+\rho^{(m)}$ can be considered as constant over the
entire domain, except for the small variations at the droplet
interface.  This approach to model turbulent emulsion has the
advantage of being simple and computationally efficient. It was however
noticed that for extremely long simulation times, as those needed in
order to collect firm statistical information, one can observe
important diffusion of one fluid component into the other. This
physical phenomenon can be artificially large in the LBM simulations
due to the relatively big interface width between the two fluid
components. We have however shown in a recent publication \cite{per11_jpcs}
how this effect can be controlled by means of an effective and
computationally efficient manner. By means of this improvement, the
presented algorithm allows us to investigate turbulent emulsions for
arbitrarily long simulations times.

In order to stir turbulence we apply forcing at each position and at
each time step modulated by means of a sum of sine waves with small
wavenumbers. To ensure an homogeneous and isotropic stirring, the
phases of the sine waves are evolved in time by means of a stochastic
process.  The forcing is divergenceless and its expression for the
generic $i$-{\em th} component is:
\begin{equation}
  \nonumber
  F^{(\alpha)}_i({\bm x},t) = A \rho^{(\alpha)} \sum_{j\neq i} \left[\sin(k_j x_j + \Phi_k^{(j)}(t)) \right]
  \nonumber
\end{equation}
 where $i,j=\{1,2,3\}$, ${\bf F}^{(\alpha)}$ is
the external force at ${\bm x}$ and time $t$.  $A$ controls the
forcing amplitude, $k_j$ are the wave-vector components and the sum is
limited to $k^2=k_1^2+k_2^2+k_3^2\leq 2$. The phases $\Phi_k^{(j)}$ are
evolved in time according to independent Ornstein-Uhlenbeck processes
with the same relaxation times $T=u_{rms}/N$ ($N$ is the linear domain
size and $u_{rms}$ was taken equal to $0.1$ to almost match with the
typical values for the large scale velocity).
\begin{table}
  \caption{\label{tab:table1}Runs parameters. For runs ${\tt N128A-128D}$ and ${\tt N512A}$, $\rho^{(m)}+\rho^{(d)}=2.4$. For run ${\tt N512B}$, $\rho^{(m)}+\rho^{(d)}=1.077$. Here $\nu$ is the viscosity of the multicomponent fluid, $G$ is the interaction strength and $R_{\lambda}$ is the Taylor's based Reynolds number~\cite{Fri96}. The relaxation time of the two fluids $\tau=0.515$ was kept fixed.}	
\begin{ruledtabular}
  \begin{tabular}{c c c c c c c c c}
    & $N$ & $G$ & $\varepsilon$ & $Re_{\lambda}$ & $\nu$ & $\sigma$ & $\phi ({\mbox{in } \%})$\\
      \hline
      \hline
      ${\tt N128A}$ & $128$ & $0.3$ & $4.0\cdot10^{-8}$ & $15$ & $5\cdot 10^{-3}$ & $1.6\cdot 10^{-3}$ & $0.07$ \\
      ${\tt N128B}$ & $128$ & $0.3$ & $4.0\cdot10^{-8}$ &$15$ & $5\cdot 10^{-3}$ & $1.6\cdot 10^{-3}$ & $0.5$ \\
      ${\tt N128C}$ & $128$ & $0.3$ & $4.0\cdot10^{-8}$ & $15$ & $5\cdot 10^{-3}$ & $1.6\cdot 10^{-3}$ & $5.0$ \\
      ${\tt N128D}$ & $128$ & $0.3$ & $4.0\cdot10^{-8}$ & $15$ & $5\cdot 10^{-3}$ & $1.6\cdot 10^{-3}$ & $10$ \\
      ${\tt N512A}$ &$512$ & $0.3$ & $2.9\cdot10^{-9}$ & $30$ & $5\cdot 10^{-3}$ & $1.6\cdot 10^{-3}$ & $0.3$  \\
      ${\tt N512B}$ &$512$ & $0.1$ & $2.9\cdot10^{-9}$ & $30$ & $5\cdot 10^{-3}$ & $1.7\cdot 10^{-3}$ & $0.3$  \\
    \end{tabular}
  \end{ruledtabular}
\end{table}

\section{The stationary state}
\begin{figure}[!ht]
\vspace{-0.38cm}
\includegraphics[width=0.34\linewidth]{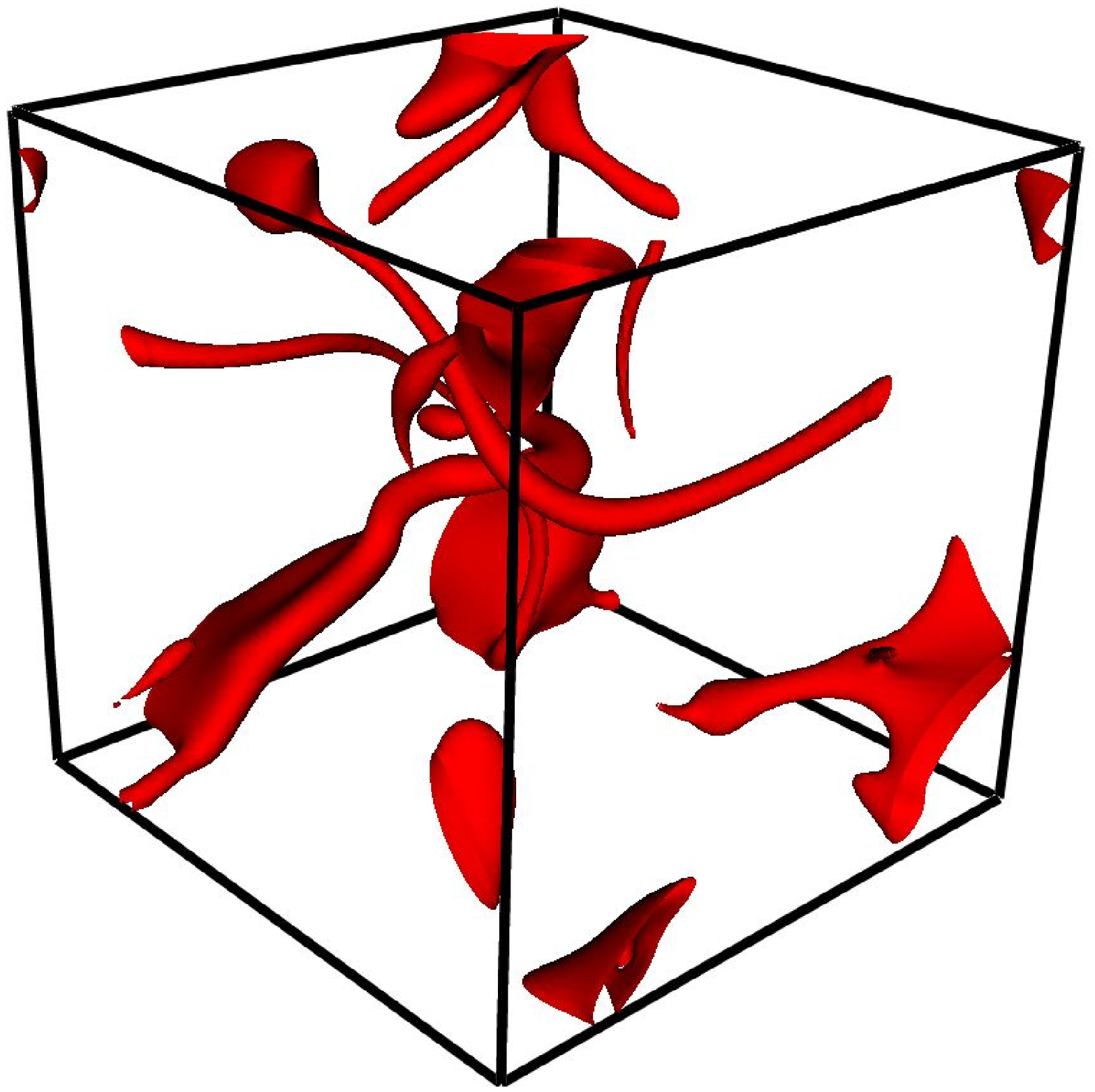}
\hspace{-0.45cm}
\includegraphics[width=0.34\linewidth]{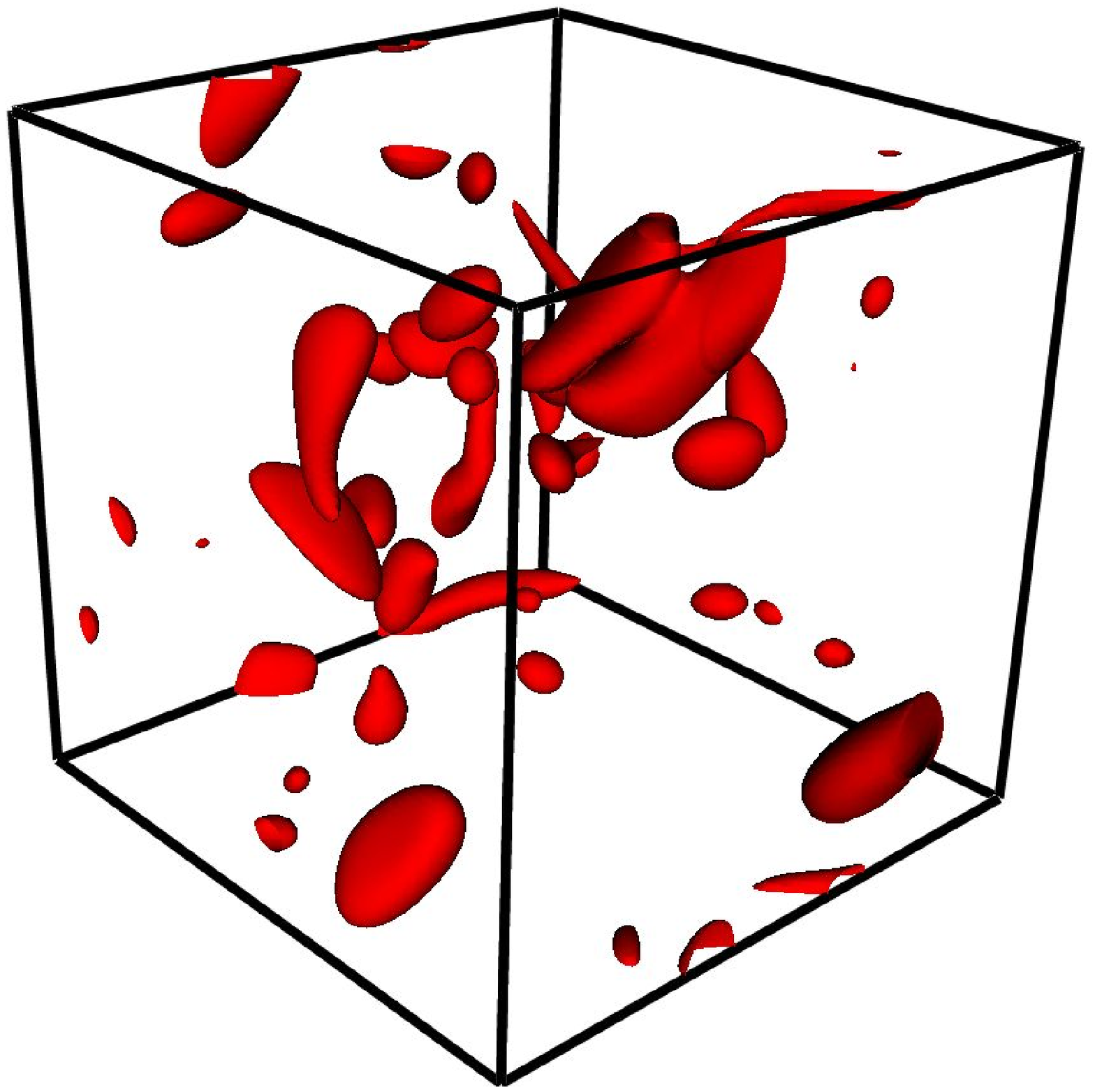}
\hspace{-0.45cm}
\includegraphics[width=0.34\linewidth]{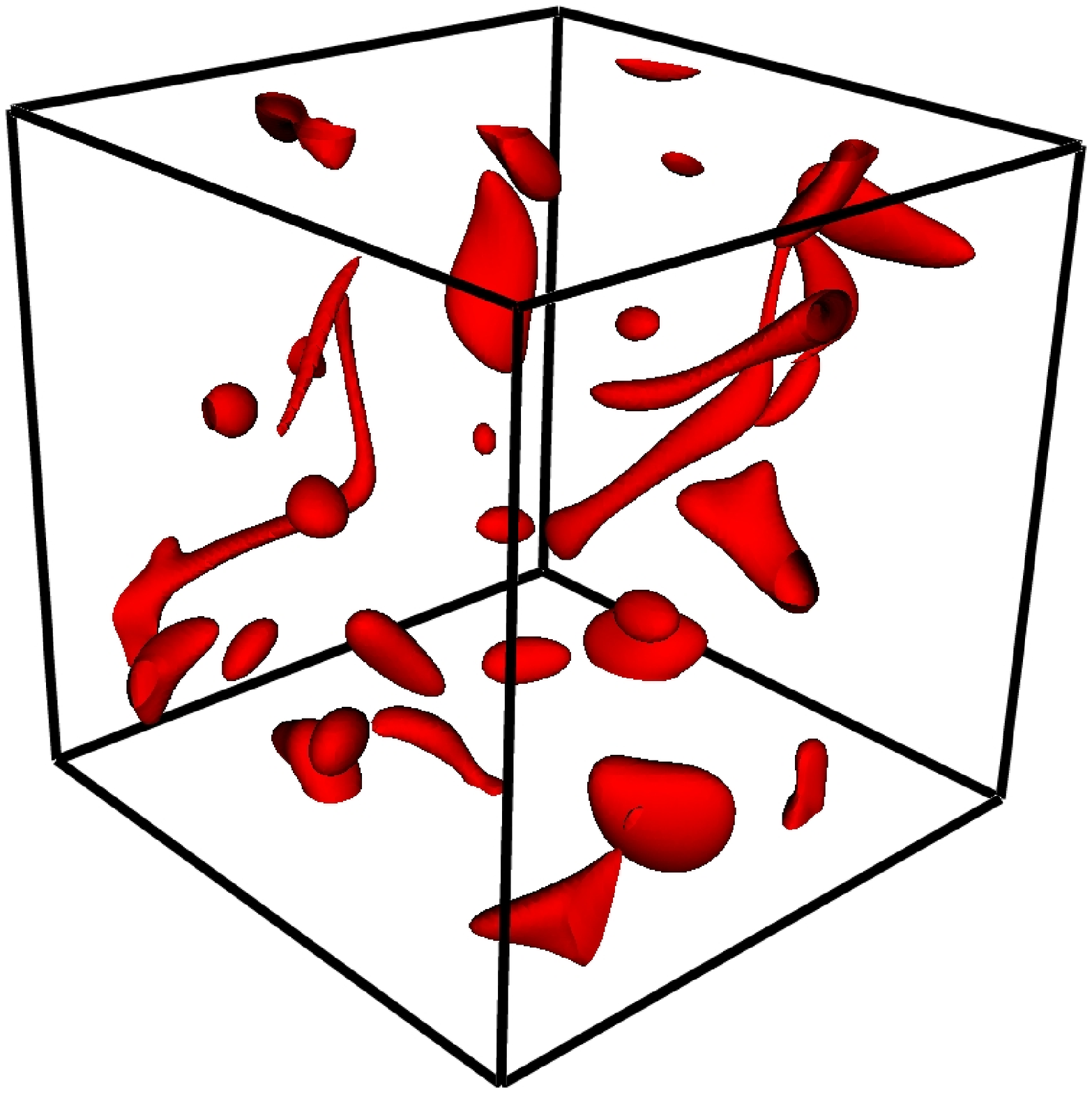}
\vspace{-0.4cm}
  \includegraphics[width=\linewidth]{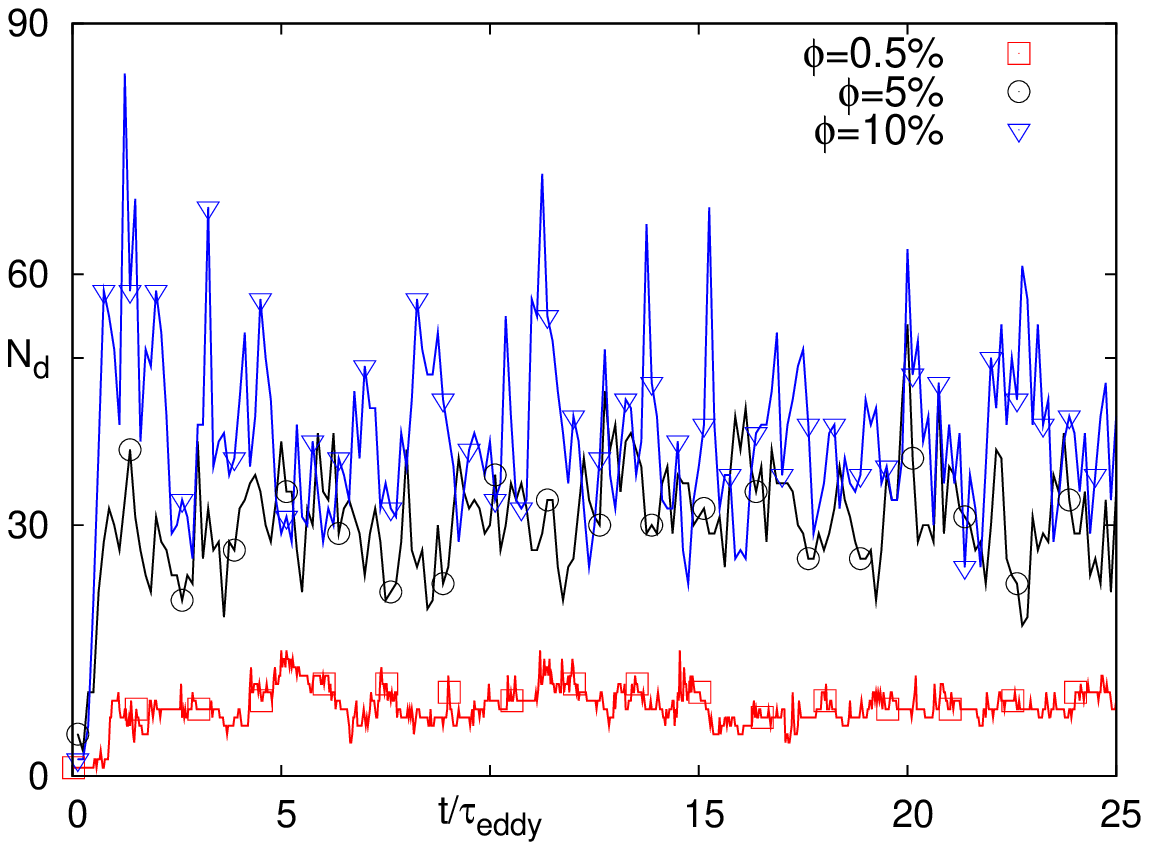}
  \caption{\label{fig:numdN128} Number of droplets vs. time for
    different values of the volume fraction, $\phi$. An increase in
    the volume fraction leads to an increase in the total droplets
    number and more coagulation processes occur. The time is made
    dimensionless by the large scale eddy turnover time,
    $\tau_{eddy}=N/u_{rms}$. Top panel shows snapshots of the 
    droplet dispersion at the intermediate $t/\tau_{eddy}=0.33$ and steady 
    state $t/\tau_{eddy}=12.5, 25$  configurations.}
\end{figure}
\begin{figure}[!ht]
  \includegraphics[width=0.8\linewidth]{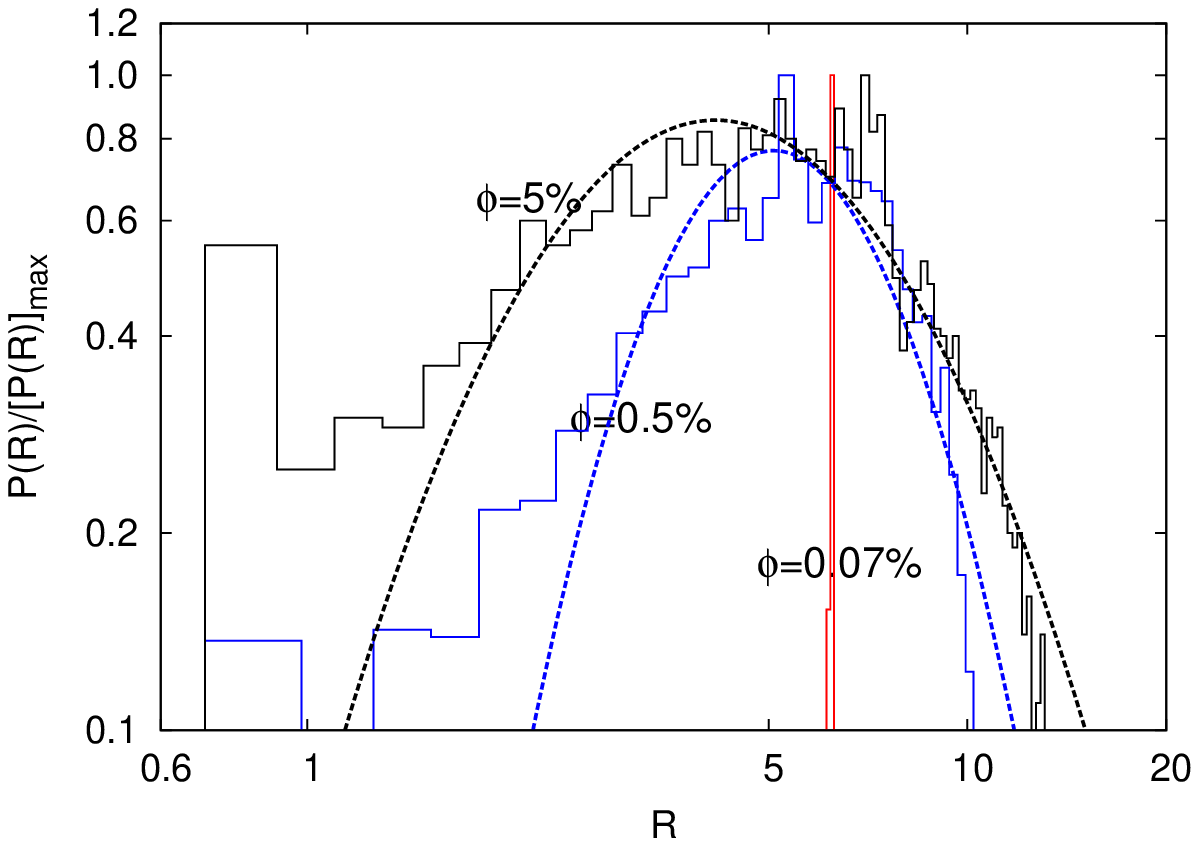}
  \includegraphics[width=0.8\linewidth]{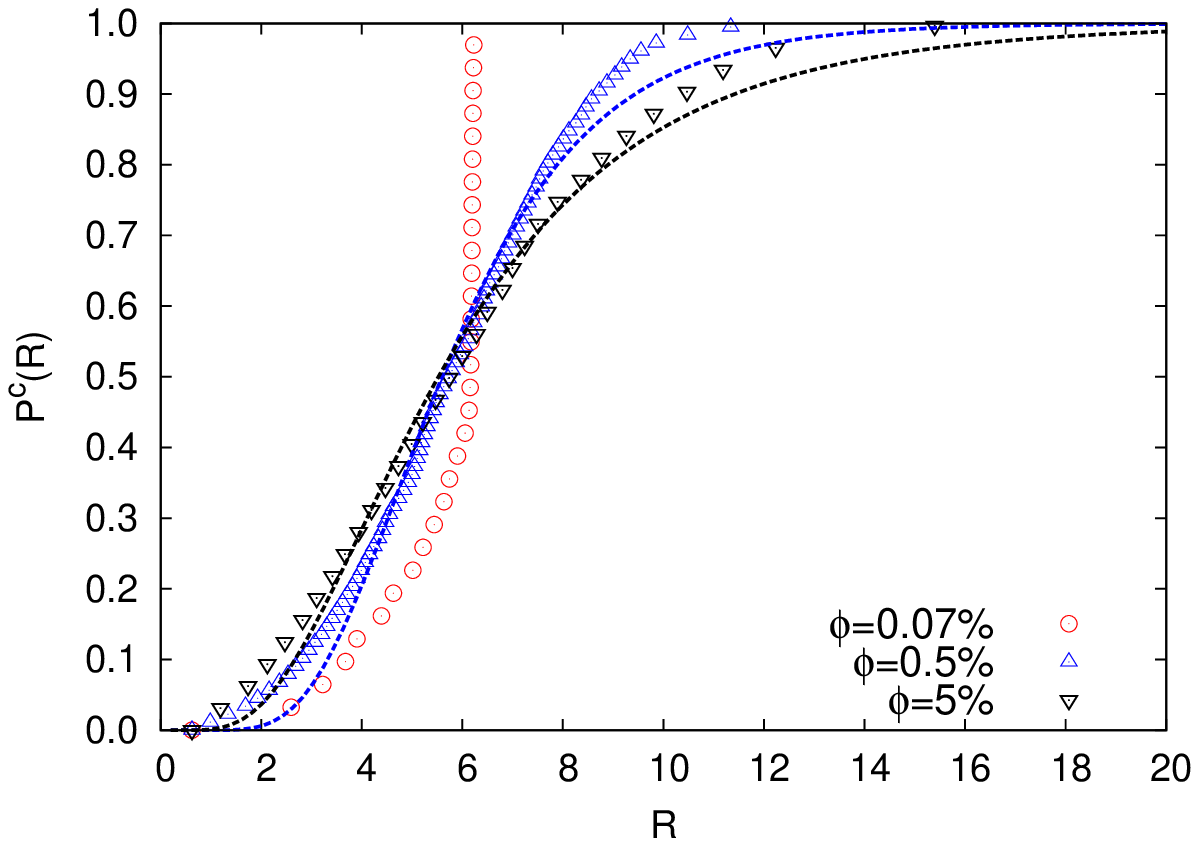}
  \caption{\label{fig:pdfrN128} (top-panel) Log-log plot of the
    probability distribution function (pdf) of droplet radius for
    different values of the volume fraction. Pdf are scaled to have
    maximum value unity. (bottom-panel) Cumulative probability
    distribution function for different volume fractions. In panels,
    dashed lines represent log-normal fits to distributions with
    $\phi=0.5\%$ and $\phi=5\%$.}
\end{figure}

In presence of turbulence droplets continuously undergo breakup and
collision processes, whose rate depends on both volume fraction and
Reynolds number. The plot in Fig.~\ref{fig:numdN128} shows a typical
time evolution of the droplet number for volume fractions $\phi$ equal
to $0.5\%$, $5\%$ and $10\%$ at $Re_{\lambda}=15$ (runs ${\tt N128B-D,
  Table~\ref{tab:table1})}$. During the early simulation stages, when
turbulence is still developing, the droplet deforms and then starts to
breaks up into smaller droplets.  At later times, the droplet count
steadily oscillates around a mean value because of a continuos
competition between breakups and coalescence events.  The stationary
probability distribution function of droplet sizes, with its
corresponding cumulative distribution, is shown in
Fig.~\ref{fig:pdfrN128} for different volume fractions $\phi=0.07\%$,
$\phi=0.5\%$ and $\phi=5\%$ with $Re_{\lambda}=15$ and
$\sigma=1.6\cdot 10^{-3}$ (runs ${\tt N128A-C}$,
Table~\ref{tab:table1}).  \REM{general comment,I do not like to call
  the run in the text with their acronym. shall we identify them with
  physical parameters (Reynolds,colume fraction etc...) }  

In the case $\phi=0.07\%$ only very few breakups are observed and the
droplet PDF closely resemble a delta function.  We observe a mild
increase in the average droplet radius at increasing the volume
fraction of the droplet phase; furthermore, because of enhanced
coalescence, we observe that the maximum droplet radius increases with
increasing the volume fraction (see
Fig.~\ref{fig:pdfrN128}(bottom)). In the same figure we also superpose
the empirical pdf with a log-normal distribution. 

As one can see, the two fits are in good agreement with the data,
within statistical errors, at large and intermediate radiuses. At very
small droplet radius one can observe deviations between measured data
and the fitting function. On this point we are not able to draw
conclusive statement and we tend to think that deviations may be
unphysical and due to the limitations of the diffused interface models
to capture droplets whose size is comparable to the interface width.

\begin{figure}[!t]
  \includegraphics[width=\linewidth]{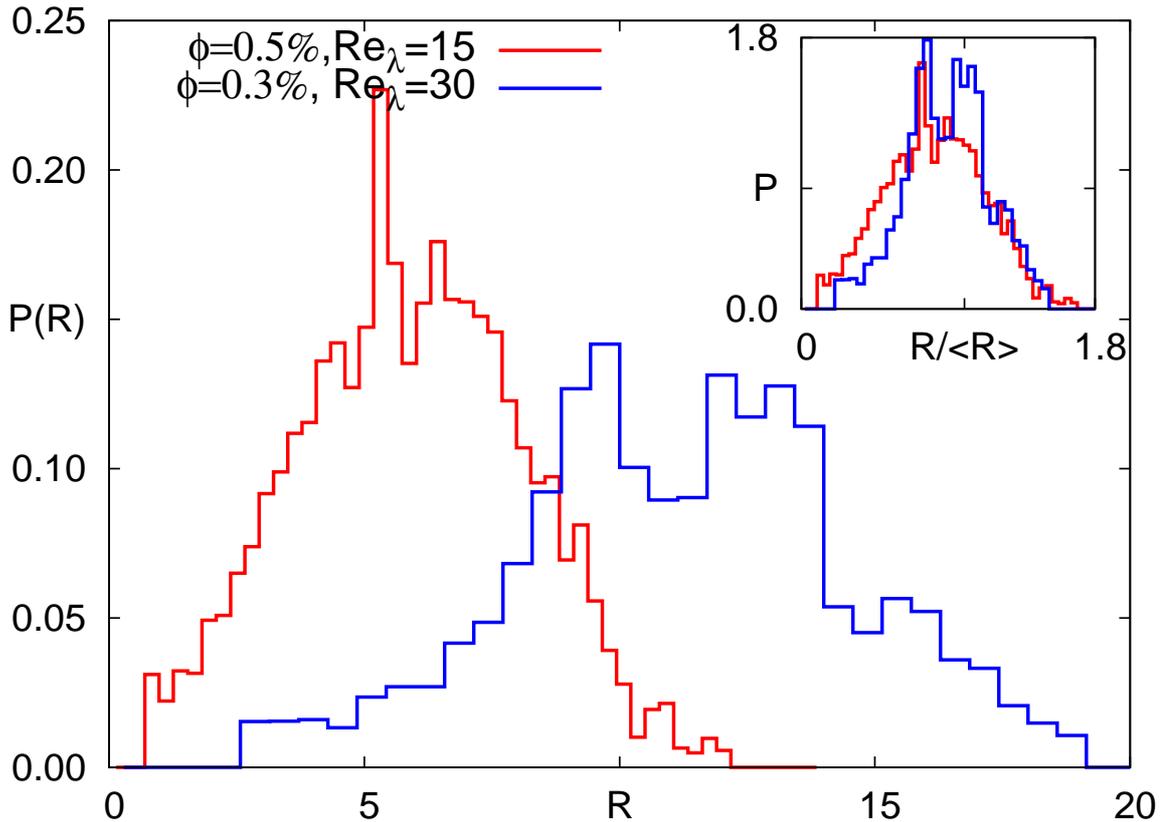}
  \caption{\label{fig:pdfrN128re} Probability distribution function of
    the droplet radius from two simulations with almost the same
    volume fraction but with different $Re_{\lambda}$. The inset shows
    the pdf normalized by the average droplet radius.}
\end{figure}

To understand the effect of the Reynolds number on the stationary
droplet size distribution, in Fig.~\ref{fig:pdfrN128re} we compare the
droplet radius PDFs with comparable volume fractions but for different
Reynolds number $\phi=0.5\%, Re_{\lambda}=15$ and $\phi=0.3\%,
Re_{\lambda}=30$ (corresponding to runs ${\tt N128B}$ and ${\tt
  N512A}$, Table~\ref{tab:table1}).  Our result support the validity
of the Hinze criterion which predicts that, for the same material
parameters, the average droplet diameter should only depend inversely
on the energy dissipation rate [Eq.~\eqref{eq:hinze}]. By rescaling
the droplet radius with their average values, the two PDF collapse
(inset of Fig.~\ref{fig:pdfrN128re}). Thus, we can exclude strong
dependency of the pdf shape, at fixed volume fraction and material
properties, on the Reynolds number; at least this is the case for the
two Reynolds number that we investigated. 
\REM{general comment,I do not like to call the run in the text with
  their acronym. shall we identify them with physical parameters
  (Reynolds,colume fraction etc...) }

\begin{figure}[!t]
  \includegraphics[width=\linewidth]{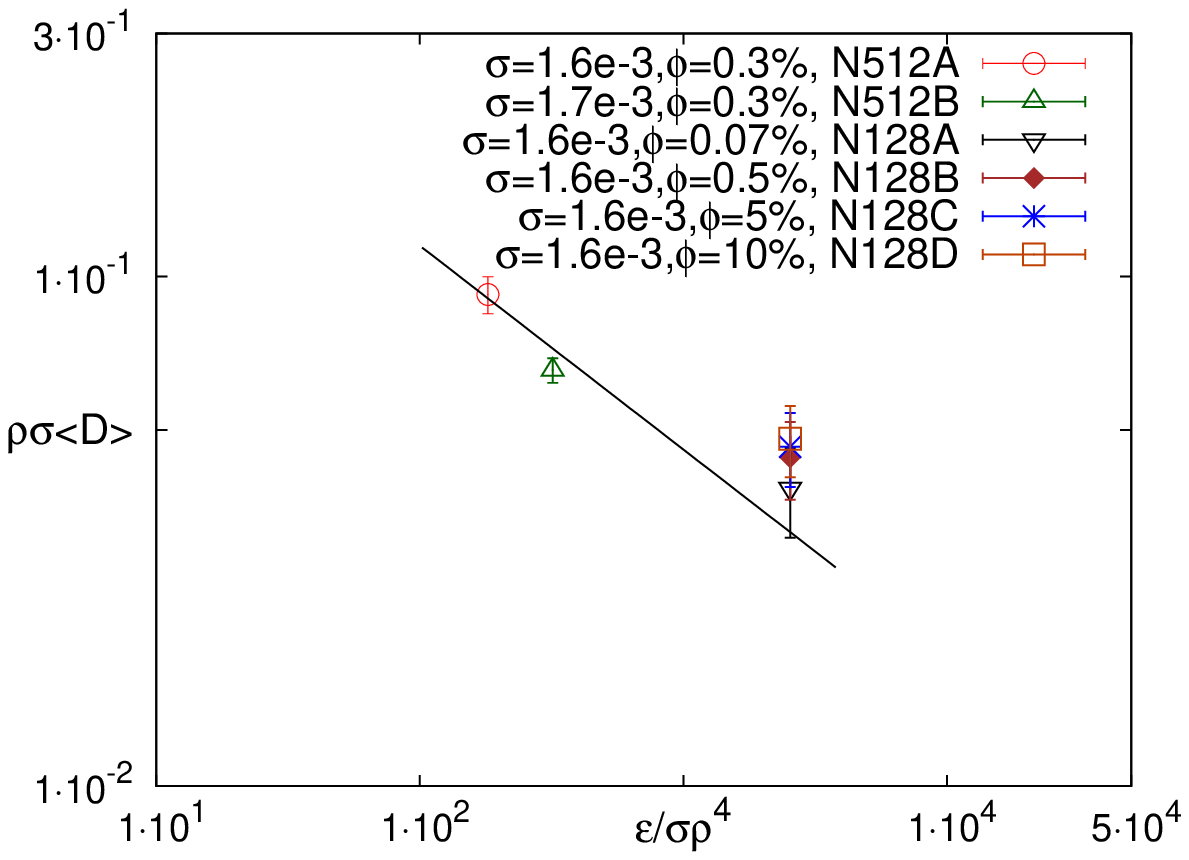}
  \caption{\label{fig:hinzp}Droplet diameter versus the energy
    dissipation rate to test the Hinze criterion for droplet breakup
    \cite{hin55}.  The different symbols correspond to runs that are
    distinguished by their volume fraction $\phi$ and surface tension
    $\sigma$ (see Table~\ref{tab:table1}).  \REM{${\tt N128A}$ (black
      inverted triangle), ${\tt N128B}$ (brown diamond), ${\tt N128C}$
      (half filled blue circle), ${\tt N128D}$ (quarter filled brown
      circle), ${\tt N512A}$ (green triangle), and ${\tt N512B}$ (red
      circle).}  The solid black line is the prediction of Hinze
    [Eq.~\eqref{eq:hinze}]. Note how at larger volume fractions
    deviations from the Hinze criterion are observed, these are
    probably due to enhancement of coagulation events. The error-bars
    indicate the uncertainty in the estimation of the droplet
    interface position.}
\end{figure}

{\section{Hinze criterion}}
The Hinze criterion, as in Eq.~\eqref{eq:hinze}, provides the estimate
for the maximum droplet diameter, $D_{max}$, that should not undergo
breakup at a fixed turbulence intensity. In presence of turbulence
however breakup and coalescence always occurs and the correct
indicator for the droplet size is the average droplet radius. We
replace therefore $D_{max}$ by the average droplet diameter $\langle D
\rangle$, where with the angular bracket we indicate averaging over a
steady distribution of droplet diameters.
From Fig. \ref{fig:hinzp} one can see that our data support the
validity of Hinze criterion in the case of small volume fractions,
while we one can clearly detect a non trivial dependency on the volume
fraction for larger values of $\phi$ (keeping the turbulent intensity
and surface tension constant).

\begin{figure}[!h]
\begin{center}
  \includegraphics[width=\linewidth]{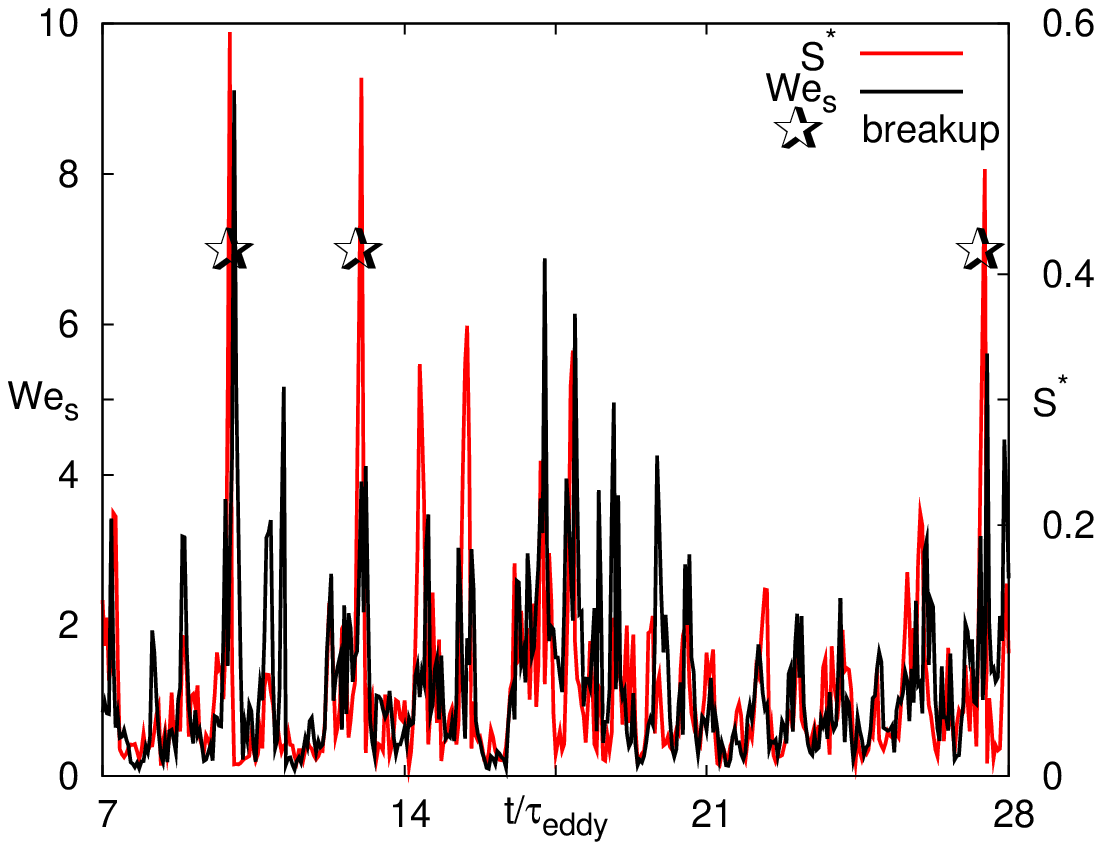}
\end{center}
\caption{\label{fig:WSs}Correlation between the Weber number and the
  non-dimensional droplet extension $S^* \equiv (S/S_0-1)$. The
  instantaneous Weber number is computed by using the maximum value of
  the energy dissipation at the droplet surface $We_s$.  
  The plot represent the evolution of a typical droplet trajectory
  from the run with smallest volume fraction $\phi=0.07\%$ and
  $Re_{\lambda}=15$ (${\tt N128A}$). Stars indicate the breakup events
  that we observed during this particular run.}
\end{figure}
\begin{figure}[!t]
  \begin{center}
    \includegraphics[width=1.\hsize]{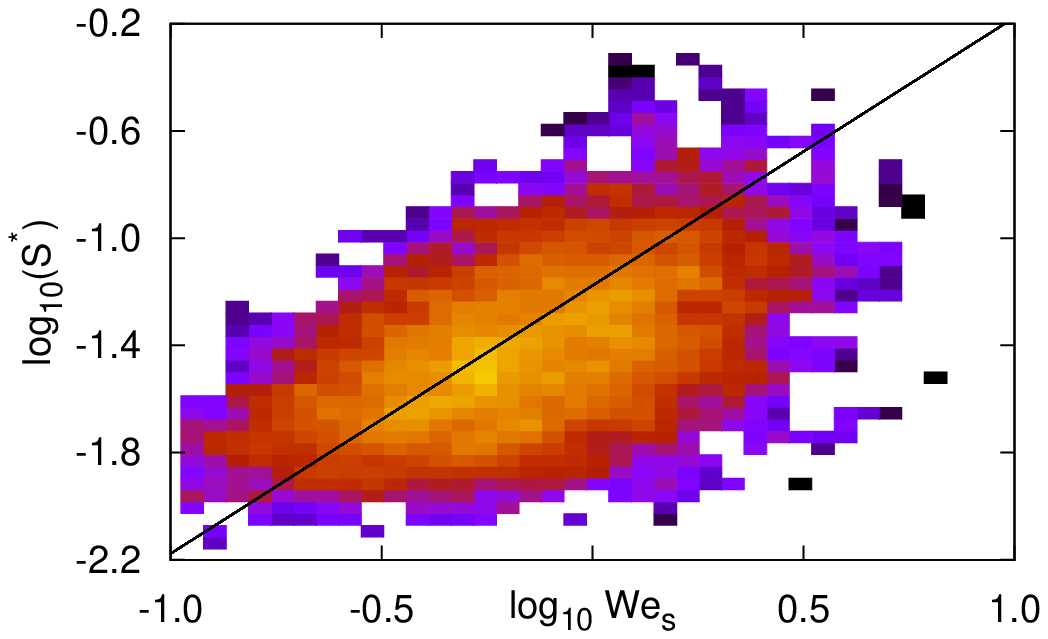}
  \end{center}
  \vskip -1cm
  \caption{\label{fig:correl}Log-log plot of the joint probability
    distribution function of the non-dimensional droplet extension
    $S^*$ versus the local Weber number, as obtained from the velocity
    at the droplet surface $We_s=|u_{max,S}-u_{CM}|$ (see text) for
    the run with $\phi=0.07\%$ and $Re_{\lambda}=15$ $({\tt
      N128A})$. From a least squares fit we obtain $S^*\approx
    {0.066}We$ (black line) which is in agreement with earlier
    simulations~\cite{qia06}. In the scatter plot the lighter central
    region (yellow in color figure) indicate the region of high
    probability while the darker (black) regions indicate low
    probability.}
\end{figure}

As the droplet is transported by the turbulent flow, it visits regions
of where velocity gradients are much larger or smaller than the
average value. Thus, a better characterization of the droplet breakup
required the tracking of individual droplets along their trajectory
along with monitoring their deformation and the local flow conditions
in the immediate neighboroud, i.e. the local Weber number,
$We_s$. Here we define the local Weber number as the difference
between the maximum velocity difference between a point on the droplet
surface and the velocity of the droplet center of mass $We_s \equiv
|{\bm u}_{max,S}-{\bm u}_{CM}|$.  The droplet deformation is instead
characterized by a dimensionless parameter $S^*=(S/S_0-1)$ where $S$
is the surface area of the droplet and
$S_0\equiv(4\pi)^{1/3}(3V_d)^{2/3}$ is the surface area of the
equivalent spherical droplet (i.e. a spherical droplet with the same
volume of the deformed one). With this definition $S^*=0$ corresponds
to an (undeformed, i.e. spherical) droplet, whereas $S^*>0$ indicate a
stretched droplet. The plot in Fig.~\ref{fig:WSs} shows the time
evolution of the surface deformation, $S^*$, of the local Weber number
$We_s$, and times at which breakup event occur, during a part of the
droplet trajectory. Data in Fig.~\ref{fig:WSs} refers to the run with
the smallest volume fraction, $\phi=0.07$, and with $Re_{\lambda}=15$.
For this run, due to its low droplet count, it was technically easier
to track breakup events (run ${\tt N128A}$,
Table~\ref{tab:table1}). As one can see, the droplet deformation,
$S^*$, correlates strongly with the local Weber number,
$We_s$. Furthermore, just before the droplet breaks (stars in the
figure) one can identify a sharp increase in both surface deformation
$S^*$ and $We_s$. We can further quantify the correlation by computing
the joint pdf of $S^*$ and $We_s$ over the full time-series for run
${\tt N128A}$ (see Fig.~\ref{fig:correl}). From this joint PDF we find
that $S^*\approx {0.066}We$. In Ref.~\cite{qia06} it was reported a
similar correlation but for the rather different case of a
vapor-liquid dispersion with a density contrast between the two
phases.

\section{Conclusions}
We have studied the droplet dispersion in a turbulent emulsion from
very small to larger volume fractions. We focused on the role of
turbulence fluctuations on the droplet deformation and breakup and to
this end we limited to density and viscosity matched emulsions. In
this way we could better highlight the new physics associated with the
interplay between surface tension and turbulence fluctuations. Our
numerical study, which is based on the lattice Boltzmann method, shows
that by using a proper large-scale isotropic stirring mechanism
\cite{per11_jpcs}, it is possible to attain a stationary steady state which
allows to study the physics of droplets distribution and evolution in
turbulence. The pdf of droplets radii follow a log-normal distribution
and the Hinze criterion is well satisfied at small volume fraction,
while we observe a departure from its prediction at higher volume
fractions. We performed a Lagrangian tracking of individual droplets
and we showed that the local Weber number is a suitable prognostic
indicator for droplet breakup and correlates strongly with droplet
deformation.

{\section{Acknowledgements}}
We thank S. Sundaresan, J. Derksen and H. Xu for discussions.  We
acknowledge the COST Action MP0806 and FOM (Stichting voor
Fundamenteel Onderzoek der Materie) for support. LB, PP, and FT
acknowledge the Kavli Institute of Theoretical Physics for
hospitality. This research was supported in part by the National
Science Foundation under Grant No. NSF PHY05-51164. We acknowledge
computational support from CASPUR (Roma, Italy), from CINECA (Bologna,
Italy), and from JSC (Juelich, Germany).

\end{document}